\documentstyle[prl,preprint,aps]{revtex}
\begin{document}
\title {Evolutionary freezing in a competitive population} 
\author{N.F. Johnson$^1$, D.J.T. Leonard$^1$, P.M. Hui$^2$ and T.S.
Lo$^2$}  
\address {$^1$ Department of Physics, University of Oxford,\\   Parks Road,
Oxford, OX1 3PU, England, United Kingdom}
\address {$^2$ Department of Physics, The Chinese University of Hong
Kong,\\ Shatin, New Territories, Hong Kong}
%
\maketitle


\begin{abstract}

We show that evolution in a population of adaptive agents, repeatedly
competing for a limited resource, can come to an abrupt halt. This
transition from evolutionary to non-evolutionary behavior arises as the
global resource level is changed, and is reminiscent of a phase transition
to a frozen state. Its origin lies in the inductive decision-making of the
agents, the limited global information that they possess and the dynamical
feedback inherent in the system. 

\vskip\baselineskip
\vskip\baselineskip
\end{abstract}


\newpage

An evolving population in which individual members (`agents') adapt their
behavior according to past experience while repeatedly competing for some
limited global resource,  is of great interest to the social, economic and
biological sciences
\cite{bak,casti,Arthur,Huberman,Sigmund,amaral}. Arthur
\cite{Arthur} has introduced a simple example, called the bar-attendance
problem, which typifies the complexity often encountered in these different
disciplines \cite{casti}: potential customers of a particular bar with a
limited seating capacity have to repeatedly decide whether or not to attend
on a given night each week
\cite{johnson1}.  A special limiting case of this problem, the so-called
minority game \cite{challet}, is currently attracting much attention 
\cite{challetmarsili,savit,rodgers} and has been shown to have a
fascinating connection with disordered spin systems
\cite{challetmarsili}.   

In this paper we introduce and study a simple implementation of the
general bar-problem which includes evolutionary learning by the attendees. 
We find that the evolutionary process can come to an abrupt halt as the
global resource level is varied. This transition is dynamical in origin: it
results from the inductive decision-making of the agents, the limited
global information that they possess, and the dynamical feedback present in
the system's memory.

Consider $N$ agents repeatedly deciding whether to attend a bar with a
seating capacity (i.e. global resource level) of $L$.  Let the actual
attendance at the bar at time-step $t$ be $A_{t}$.  If $A_{t} \leq L$ , the
outcome is the signal  `undercrowded'. In contrast, if
$A_{t} > L$ then the outcome is the signal `overcrowded'.  Hence, the 
outcome can be represented by a binary string of $0$s (representing, say,
`undercrowded') and $1$s (representing `overcrowded'). The outcome is the
{\em only} information made known to all agents: the value $L$ is not
announced, and the agents cannot communicate with each other. All agents
have the same level of capability: specifically, each agent has access to a
common register or `memory' containing the outcomes from the most recent 
occurrences of all $2^m$ possible bit-strings $\{\alpha\}$ of length $m$. 
This register can hence be written at timestep $t$ as a
$2^m$-dimensional vector
$\bf h_t$ with binary components
$h^\alpha_t\ \in
\{0,1\}$ corresponding to the outcome of the most recent occurrence of
history $\alpha$: we will call $\bf h_t$ the `predicted trend' at time
$t$. We assign each agent a single strategy
$p$. Following a given
$m$-bit sequence $\beta$,
$p$ is the probability that the agent will choose the same outcome
$h^{\beta}_t$ as that stored in the common register, i.e. she follows the
predicted trend; 
$1-p$ is the probability she will choose the opposite, i.e.  reject the
predicted trend.

The `good' decisions are attending (not attending) the bar with the outcome
being `undercrowded'  (`overcrowded'). The `bad' decisions are attending
(not attending) the bar with the outcome being `overcrowded'
(`undercrowded').   After the outcome at a given time $t$ is announced, the
agents making `good' (`bad') decisions gain (lose) one point. If an agent's
score  falls below a value
$d<0$, then her strategy is modified, i.e. the agent gets a new $p$ value
which is chosen with an equal probability from a range of values, centered
on the old
$p$, with a width equal to $R$.  Upon strategy modification, the agent's
score is reset to zero. Hence $d$ is the net number of times an agent is
willing to be wrong before modifying her strategy. Although this
evolutionary procedure provides a fairly crude `learning' rule as far as
machines are concerned
\cite{Sutton}, in our experience it is not too dissimilar from the way that
humans actually behave in practice.  Changing
$R$ allows the way in which the agents learn to be varied. For $R=0$, the
strategies will never change (though the memory register will).  If $R=2$,
the strategies before and after modification are uncorrelated. For small
$R$, the new
$p$ value is close to the old one. Our results are insensitive to the
particular choice of boundary conditions employed.  For $L=(N-1)/2$,  the
model reduces to the evolutionary minority game introduced in Ref. 
\cite{ourPRL}.

Figure 1 shows the numerical results for the mean and standard deviation of
the bar attendance as a function of the global resource level (i.e. seating
capacity) defined as the percentage
$\ell=(100\times  L/N)\ \%$.  Here 
$N=1001$,
$m=3$, $d=-4$ and $R=2$ although the same general features arise for other
parameter values (see Ref. \cite{uspreprint}). Averages are taken over
$10^4$ timesteps within a
given run, and then over 10 separate runs with random initial
conditions: this average is denoted by
$\langle \cdots
\rangle$. Here we will focus on the mean attendance 
$\langle A \rangle$; the standard deviation of the attendance
$\Delta A$ given by $[\langle A^2\rangle -  \langle
A\rangle^2]^{\frac{1}{2}}$; the average predicted trend given by 
$\langle h^\alpha\rangle$; the average strategy 
$\langle p\rangle$ given by
$\frac{1}{N}\int_0^1 p P(p) dp$ where $P(p)$ is the strategy
distribution among the agents. $P(p)$ satisfies $\int_0^1 P(p) dp = N$
and is time- and run-averaged as discussed above.  From the definition
of the game, games with cutoff
$L'
\equiv N-L$ are related to games with cutoff $L$ in that the mean
attendance in the game with
$L'$ can be found from the mean population of agents {\em not} attending
the bar in a game with $L$. Hence we focus on $L \geq N/2$. The resulting
symmetry about $\ell=50\%$ is clearly shown in Fig. 1.

At $\ell=50\%$, the mean attendance $\langle A\rangle=N/2$ while $\Delta A$
is {\em smaller} than for the `random' game in which agents independently
decide by tossing a coin ($\Delta A_{\rm random}\approx {\sqrt N}/2
=15.8$). As $L$ and hence
$\ell$ increase ($\ell>50\%$), the mean attendance $\langle A\rangle$
initially shows a small plateau-like structure, then increases steadily
while always lying below the value of $L$. This increase in
$\langle A\rangle$ occurs despite the fact that the value
$L$ (and hence $\ell$) is unknown to the agents.  The standard deviation
$\Delta A$ increases rapidly despite the fact that the number of
available seats in the bar is actually increasing. It seems that by
increasing the level of available resources, the system actually becomes
less efficient in accessing this resource. Most surprising, however, is the
abrupt transition in both the mean attendance and standard deviation which
occurs around
$\ell_c=75\%$. (The precise value of $\ell_c$ depends on the strategy
modification range $R$ \cite{uspreprint}). For
$\ell>\ell_c$, both the mean attendance and standard deviation become
constant, regardless of the increase in the seating capacity $L$. In this
region, the bar is practically always undercrowded since $(\langle A\rangle
+\Delta A)\ll L$,  thereby offering a significant `arbitrage' opportunity
for the selling of `guaranteed' seating in advance. 

Figure 2 shows the corresponding variation in the average predicted
trend 
$\langle h^\alpha\rangle$ and the average strategy 
$\langle p\rangle$, as a function of the global resource level $\ell$.
Remarkably, 
$\langle h^\alpha\rangle$ has a step-like structure with abrupt changes at
$\ell_c$ and $\ell=50\%$. These abrupt transitions are a direct result of
the dynamical feedback and memory in the system. A random
$\bf h_t$ generator as provided by an exogenous source -- e.g.  sun-spot
activity  for prediction of financial markets -- will not contain these
discrete steps and hence does not reproduce the corresponding abrupt
transitions shown in Figs. 1 and 2. For example a coin-toss  would yield
$\langle h^\alpha\rangle=1/2$ for all $\ell$; however the
endogenously-produced predicted trend in the present model only takes this
value at {\em exactly}
$\ell=50\%$. 

The strategy distribution $P(p)$ is symmetric at
$\ell=50\%$, with peaks around $p=0$ and $p=1$ \cite{ourPRL}\cite{rodgers}.
As
$\ell$ increases, $P(p)$ becomes asymmetric. The peak around
$p=1$ becomes larger while that around $p=0$ becomes smaller: agents hence
tend to follow the predicted trend $h^\alpha_t$ more often than not.
However, an abrupt change occurs at
$\ell=\ell_c$. Figure 3 (upper) shows
$P(p)$ just either side of $\ell_c$, at 
$\ell=\ell_c\pm 2\%$. For
$\ell=\ell_c-2\%$ (solid line) the asymmetric distribution is clear but
$P(p)$ remains non-zero for all
$p$. For
$\ell=\ell_c+2\%$ (dashed line),
$P(p)=0$ for $p<1/2$ but has an irregular shape for $p>1/2$. The
final form $P(p)$ depends on the initial strategy distributions
at time $t=0$ and the strategy modification range $R$. For
$\ell_c<\ell\leq 100\%$,  the distribution $P(p)$ and hence $\langle
p\rangle$ remain essentially unchanged.

Figure 4 shows the average lifespan per agent defined as 
$\tau=\frac{1}{N}\int_0^1 P(p) \Lambda(p) dp$, where $\Lambda(p)$ is the
average number of turns between strategy-modifications experienced by an
agent at $p$. As
$\ell\rightarrow
\ell_c$, this lifetime $\tau$ becomes effectively infinite (i.e. the
duration of the simulation) and remains at this value for all
$\ell>\ell_c$. Above
$\ell_c$ all agents have a strategy value $p>1/2$: the agents do not modify
their strategies and hence  keep this same $p$ value indefinitely. There is
hence no evolution in this system  for $\ell>\ell_c$, despite the fact that
the bar has increased seating capacity which could be exploited (Fig. 1).
In essence, the agents are winning enough points to avoid
strategy-modification: although the population as a whole is not acting
optimally, the system chooses not to employ any additional evolution. The
frozen, non-optimal steady state has a huge degeneracy. These findings
suggest to us the following informal analogy with a magnetic system  with
$0\leq p<1/2$ representing spin-down and $1/2< p\leq 1$ representing
spin-up. The point
$\ell=50\%$ corresponds to an antiferromagnet at zero field: since $P(p)$
is symmetric, there is no net magnetization. As the external `field'
$(\ell-50)\%$ increases, there is some weak paramagnetism yielding a small
value of the magnetization because
$P(p)$ is now biased towards $p=1$ (Fig. 3 upper graph, solid line). At the
critical field
$(\ell_c-50)\%$, there is a transition to a ferromagnetic state (i.e. pure
spin-up with $p>1/2$) as a result of the change in $P(p)$ shown in Fig. 3.

A connection between the behaviors of the `macroscopic' variables in Fig. 1
and the `microscopic' variables in Fig. 2, can be made analytically. It can
be shown \cite{uspreprint} that the mean attendance 
\begin{equation}
\langle A\rangle \approx N [\langle h^\alpha\rangle + \langle p\rangle -2
\langle h^\alpha\rangle \langle p\rangle]
\end{equation} and the standard deviation 
\begin{equation}
\Delta A \approx 2 N [\langle h^\alpha\rangle (1-\langle
h^\alpha\rangle)]^{\frac{1}{2}} |\langle p\rangle - \frac{1}{2}| \ .
\end{equation} Here $\langle h^\alpha\rangle$ and $\langle p\rangle$ are
implicitly functions of $\ell$. At $\ell=50\%$,
$\langle h^\alpha\rangle=1/2$ and $\langle p\rangle=1/2$ (Fig. 2) hence
$\langle A\rangle\approx N/2$ in agreement with Fig. 1. For
$50\%<\ell<\ell_c$,
$\langle h^\alpha\rangle\approx 1/4$:   the ratio of  the slopes of
$\langle A\rangle$ and $N \langle p\rangle$ in Figs. 1 and 2 is
$0.50$ which agrees exactly with the analytic result obtained by
differentiating Eq. (1) with respect to $\langle p\rangle$.  Similarly, the
ratio of the slopes of
$\Delta A$ and $N \langle p\rangle$ in Figs. 1 and 2 is $0.87$ which agrees
exactly with the analytic result obtained by differentiating Eq. (2) with
respect to $\langle p\rangle$.  For $\ell>\ell_c$, $\langle
h^\alpha\rangle=0$ hence Eqs. (1) and (2) yield
$\langle A \rangle\approx N \langle p\rangle$ and $\Delta A\approx 0$, which
is again in good agreement with the numerical results of Figs. 1 and 2. 
Note that the result $\langle h^\alpha\rangle=0$ implies that the bar is
always predicted to be undercrowded: clearly this is consistent with the
outcome since the mean $\langle A\rangle$ is well below $L$ and the
standard deviation is very small. Hence we can see why the undercrowded
state provides an attractor for the game dynamics above the freezing
transition at $\ell_c$. We note in passing that the evolutionary freezing
which arises for our generalized bar model (Figs. 1-4) is qualitatively
different from the freezing described by Challet and Marsili for the
(non-evolutionary) minority game
\cite{challetmarsili}.

In summary, we have reported transitions involving an evolving  population
of adaptive agents who repeatedly compete for a limited, but adjustable,
global resource. Abrupt changes arise for both microscopic and macroscopic
variables of the system, as the level of available resources is varied. The
present results depend crucially for their existence on the dynamical
feedback and non-local time correlations present in the system.

\vskip\baselineskip

\newpage

\centerline{\bf Figure Captions}

\bigskip

\noindent Figure 1: Numerical results for the mean bar attendance $\langle
A\rangle$ and standard deviation $\Delta A$ as a function of the global
resource level (i.e. seating capacity) defined as the percentage
$\ell=(100\times L/N)\ \%$.  The parameters (see text) are
$N=1001$,
$m=3$, $d=-4$ and $R=2$. 

\noindent Figure 2: Numerical results for the average predicted trend 
$\langle h^\alpha\rangle$ and the average strategy 
$\langle p\rangle$ as a function of the  global resource level $\ell$.

\noindent Figure 3: Numerical results for the strategy distribution
$P(p)$ (upper graph) and $\Lambda(p)$ (lower graph -- see text for
definition) near the transition point
$\ell=\ell_c$. Solid line: 
$\ell=\ell_c-2\%$. Dashed line: $\ell=\ell_c+2\%$. 

\noindent Figure 4: Numerical results for the average lifespan per agent
$\tau$ as a function of the global resource level $\ell$.

\end{document}